\documentclass[printer]{aa}
\usepackage{graphicx}
\usepackage{natbib}
\begin{document}

\title{Shallow decay phase of GRB X-ray afterglows from relativistic wind bubbles}

\author{Y.W. Yu$^{1,2}$ \and Z.G. Dai$^1$}

\institute{$^1$Department of Astronomy, Nanjing University, Nanjing
210093, China
\\$^2$Institute of Astrophysics, Huazhong Normal
University, Wuhan 430079, China}\mail{Y.W. Yu (yuyw@nju.edu.cn) and
Z.G. Dai (dzg@nju.edu.cn)}

\date{Received ... / Accepted ...}
\abstract
{}
{The postburst object of a GRB is likely to be a highly magnetized, rapidly rotating compact object (e.g., a
millisecond magnetar), which could produce an ultrarelativistic electron-positron-pair wind. The interaction of
such a wind with an outwardly expanding fireball ejected during the burst leads to a relativistic wind bubble
(RWB). We investigate the properties of RWBs and use this model to explain the shallow decay phase of the early
X-ray afterglows observed by Swift.}
{We numerically calculate the dynamics and radiative properties of RWBs.}
{We find that RWBs can fall into two types: forward-shock-dominated and reverse-shock-dominated bubbles. Their
radiation during a period of $\sim 10^{2}-10^{5}$ seconds is dominated by the shocked medium and the shocked
wind, respectively, based on different magnetic energy fractions of the shocked materials. For both types, the
resulting light curves always have a shallow decay phase, as discovered by Swift. In addition, we provide an
example fit to the X-ray afterglows of GRB 060813 and GRB 060814 and show that they could be produced by
forward-shock-dominated and reverse-shock-dominated bubbles, respectively. This implies that, for some early
afterglows (e.g., GRB 060814), the long-lasting reverse shock emission is strong enough to explain their shallow
decay phase.}
{}

\titlerunning{Shallow decay phase of GRB X-ray afterglows from RWBs}
\authorrunning{Yu \& Dai}

\maketitle

\section*{1. Introduction}
One of the most puzzling features of the early X-ray afterglow light curves of the gamma-ray bursts (GRBs)
discovered by Swift (Gehrels et al. 2004) is the existence of a flattening segment (temporal indices
$\alpha\sim[0,-0.8]$), which lasts from a few hundred seconds to a few hours (Campana et al. 2005; Vaughan et
al. 2005; Cusumano et al. 2005; Nousek et al. 2006; O'Brien et al. 2006; de Pasquale et al. 2006; Willingale et
al. 2006). This feature has been widely understood as due to a long-lasting energy injection. Two kinds of
energy injection have been proposed. One kind is the so-called refreshed shock scenario with a smooth
distribution of the Lorentz factors of the ejected shells (Rees \& M\'esz\'aros 1998). The other kind, in focus
here, involves a central engine activity extending over a long period (Dai \& Lu 1998a, 1998b; Zhang \&
M\'esz\'aros 2001; Wang \& Dai 2001; Dai 2004, D04 hereafter; Zhang et al. 2006; Fan \& Xu 2006).
\par
The popular models for the origin of long and short GRBs are the collapse of a massive star and merger of a
compact binary, respectively (for recent reviews see Woosley \& Bloom 2006; Nakar 2007). These models predict
that, after a GRB, the remaining compact object seems to be a millisecond-period pulsar or a rapidly-rotating
black hole. If the pulsar is strongly magnetized (e.g., a magnetar) or the black hole has an accretion disk
lasting for a long time, these compact objects will continuously release their rotational energy through some
magnetically-driven processes to produce an energy outflow. As this magnetically-driven outflow catches up to
and then interacts with the relativistic fireball ejected during the GRB, the fireball's energy will increase.
\par
Based on this consideration, Dai \& Lu (1998a, 1998b) and Zhang \& M\'esz\'aros (2001) propose an
energy-injection model for GRB afterglows with an assumption that the energy outflow is purely composed of
low-frequency electromagnetic (EM) waves radiated by a postburst magnetar. Using this pure EM energy-injection
(PEMI hereafter) model, Fan \& Xu (2006) successfully explain the shallow decay phase of the X-ray afterglow of
GRB 051221a. However, the numerical calculations (Wang \& Dai 2001; Fan \& Xu 2006) also show that an actual
flattening segment of a light curve is steeper than an analytical estimation when the dynamics is numerically
described and the equal-arrival surface effect is considered. Thus, it could be difficult to use the PEMI model
to explain some GRB X-ray afterglows with a very flat plateau, such as GRB 060814.
\par
On the other hand, the PEMI model does not consider possible evolution of the energy outflow with radius.
Because the fluctuating component of the magnetic field in the outflow can in principle be dissipated by
magnetic reconnection and used to accelerate an associated electron-positron plasma, the outflow should
eventually become a kinetic-energy flow carried by the accelerated $\rm e^{\pm}$ pairs, even though it is
dominated by the EM energy at small radii (Coroniti 1990; Michel 1994; Kirk \& Skj{\ae}raasen 2003). This
transformation of the outflow is estimated to occur around the radius $\sim 10^7r_{\rm L}-10^9r_{\rm L}$, that
is, below the deceleration radius of the fireball ejected during the burst, where $r_{\rm L}$ ($\sim10^7\rm cm$
for millisecond magnetars) is the light-cylinder radius of the pulsar. With these arguments, D04 suggests that
it is likely to be an ultrarelativistic $e^\pm$-pair wind (but not pure EM waves) that interacts with the
fireball to influence the GRB afterglow. As a result of the interaction, a relativistic wind bubble (RWB
hereafter) is produced, which is a relativistic version of the Crab nebula.
\par
In this paper, we calculate numerically the dynamic evolution and the corresponding radiation of a RWB driven by
a millisecond magnetar by considering the inverse-Compton scattering effect and the equal-arrival surface effect
of the RWB. Because of these effects, our numerical results are different from the analytical ones of D04. In
particular, we find that RWBs fall into two types, which are dependent of the magnetic energy fractions of
shocked materials. As an example, we fit the observed X-ray afterglows of GRB 060813 and GRB 060814, which are
found to belong to two types of RWBs. We emphasize that, for some early afterglows (e.g., GRB 060814), the
long-lasting reverse shock emission is strong enough to explain their shallow decay phase, which cannot be
simulated by using the PEMI model.

\section*{2. Dynamics of a RWB}
Some energy-source models (for a review see Zhang \& M\'esz\'aros 2004) suggest that the central engine of a GRB
is a millisecond magnetar. In such models, a GRB itself may be due to neutrino and/or magnetic processes of a
rapidly-rotating, strongly-magnetized pulsar, which may eject a few shells. Collisions between the shells may
produce internal shocks, which give rise to a few pulses of an observed GRB during the prompt emission. After
the GRB, this pulsar will be braked by magnetic dipole radiation. As a result, the release of the stellar
rotational energy (at a rate of $\dot{E}_{\rm rot}$) drives an outflow, whose luminosity is estimated by
(Shapiro \& Teukolsky 1983)
\begin{equation}
\begin{array}{ll}
 L_{\rm w}=-\dot{E}_{\rm rot}\\
~~~~\simeq4.0\times10^{47}B_{\perp,14}^2R_{\rm s,6}^{6}P_{0,\rm
ms}^{-4}\left(1+{t\over T_{\rm sd}}\right)^{-2}~\rm
erg~s^{-1},\label{Lw}
\end{array}
\end{equation}
where $B_{\perp,14}=B_{\rm s,14}\sin \theta$, $\theta$ is the angle between the magnetic and rotational axes,
and $B_{\rm s,14}=B_{\rm s}/10^{14}$G, $R_{\rm s,6}=R_{\rm s}/10^6 \rm cm$, and $P_{0,\rm ms}=P_0/1$ms are the
surface magnetic field strength, the radius, and the initial period of the magnetar, respectively. The time $t$
is measured in the observer's frame. The characteristic spin-down time is $T_{\rm
sd}\simeq5.0\times10^4(1+z)B_{\perp,14}^{-2}I_{45}R_{\rm s,6}^{-6}P_{0,\rm ms}^2~\rm s$, where
$I_{45}=I/10^{45}\rm ~g~cm^2$ is the stellar moment of inertia and $z$ the cosmological redshift. In the
following calculations, the typical values of $R_{\rm s,6}$, $P_{0,\rm ms}$, and $I_{45}$ are all taken to be
unity.
%
%
\begin{figure}
\resizebox{\hsize}{!}{\includegraphics{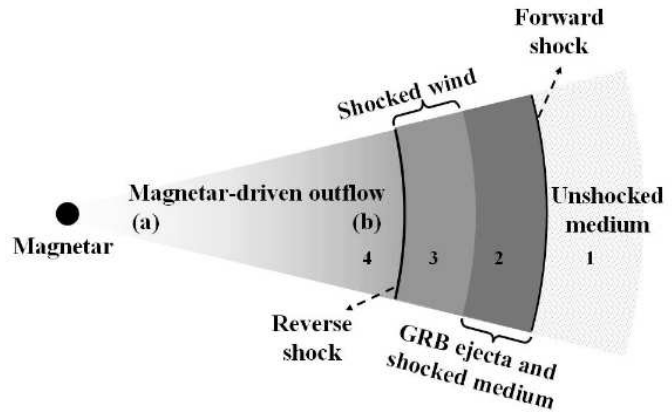}} \caption{A schematic cartoon of an RWB. The meaning of
the four regions (1-4) is explained in the text. In the inner (a) part of region 4, the energy outflow is
dominated by low-frequency EM waves, in the outer (b) part of region 4 by a kinetic-energy flow carried by
ultrarelativistic electron-positron pairs.}
\end{figure}
%
\par
As the outflow propagates outward, the energy dissipation of the EM-wave component significantly accelerates
plenty of associated $e^{\pm}$ pairs. Eventually, the outflow becomes an ultrarelativistic $e^{\pm}$-pair wind
around some certain radius that is below the deceleration radius of the shells ejected during the burst. The
wind's bulk Lorentz factor is $\gamma_{\rm w}\sim 10^4-10^7$, as argued by Atoyan (1999) for the Crab pulsar. In
the following calculations $\gamma_{\rm w}=10^4$. However, we find that a higher value of $\gamma_{\rm w}$ does
not change our results significantly. As a result of interaction between the wind and the medium, an RWB should
include two shocks: a reverse shock that propagates into the cold wind and a forward shock that propagates into
the ambient medium. Here, for simplicity, we assume that two initially-forming forward shocks during the
interaction of the GRB ejecta, both with the medium and with the wind, have eventually merged to one forward
shock. Therefore, as illustrated in Fig. 1, there are four regions separated in the bubble by the shocks: (1)
the unshocked medium, (2) the forward-shocked medium, (3) the reverse-shocked wind gas, and (4) the unshocked
cold wind, where regions 2 and 3 are separated by a contact discontinuity.
\par
We denote some quantities of region $i$ as follows: $n_i$ is the particle (proton or electron) number density
and $P_i$ the pressure measured in its own rest frame, and $\gamma_i$ and $\beta_i=(1-\gamma_i^{-2})^{1/2}$ are
the bulk Lorentz factor and corresponding velocity measured in the local medium's rest frame, respectively. The
total kinetic energy of region 2 is $E_{\rm K,2}=\gamma_2(m_{\rm ej}+m_{\rm sw})c^2+\gamma_2(\gamma_2-1)m_{\rm
sw}c^2$, where $m_{\rm ej}$ is the rest mass of the initial GRB ejecta and $m_{\rm sw}$ is the rest mass of the
swept-up medium. Energy conservation requires that any increase in $E_{\rm K,2}$ should be equal to work done by
region 3: $dE_{\rm K,2}=\delta W=4\pi R^2P_3dR$, where $R$ is the radius of the bubble in the thin shell
approximation. Then, we can obtain
\begin{equation}
{d\gamma_2\over dR}={4\pi R^2\left[P_3/c^2-(\gamma_2^2-1)n_1m_{\rm p}\right]\over m_{\rm ej}+2\gamma_2m_{\rm
sw}}.
\end{equation}
On the other hand, the dynamic evolution of region 3 can be determined by the relationship between the Lorentz
factors of the two sides of the contact discontinuity surface according to Blandford \& McKee (1976)
\begin{equation}
\gamma_3=\gamma_2\chi^{-1/2},
\end{equation}
where the similarity variable $\chi$ is (D04)
\begin{equation}
\chi=\left({L_{\rm w}\over16\pi n_1m_{\rm p}c^3\gamma_2^4R^2}\right)^{-12/29},
\end{equation}
where $m_{\rm p}$ is the proton rest mass. Using Eqs. 2, 3, 4, $P_3=L_{\rm w}/12\pi R^2\gamma_3^2c$ (D04), and
the relations of $dt=(1+z)(1-\beta_3)dR/\beta_3c$ for region 3 and $dt=(1+z)(1-\beta_2)dR/\beta_2c$ for region
2, we can get the dynamic evolution of the RWB before the characteristic time $T_{\rm sd}$. Simultaneously, the
masses of the shocked medium and shocked wind gas are calculated respectively by
\begin{equation}
{dm_{\rm sw}\over dR}=4\pi R^2 n_1m_{\rm p},
\end{equation}
\begin{equation}
{dm_3\over dR}=4\pi R^2(\beta_4-\beta_{\rm RS})\gamma_4n_4m_{\rm e},
\end{equation}
where $m_{\rm e}$ is the electron rest mass and $\beta_{\rm
RS}=(\gamma_3n_3\beta_3-\gamma_4n_4\beta_4)/(\gamma_3n_3-\gamma_4n_4)$ is the velocity of the reverse shock
measured in the local medium's rest frame. However, when $t>T_{\rm sd}$, the reverse shock is regarded as
terminative (D04) and thus Eq. 3 should be replaced by $\gamma_3\propto R^{-7/2}$ (Kobayashi \& Sari 2000;
Kobayashi 2000), as well as $n_3\propto R^{-13/2}, ~P_3\propto R^{-26/3}$, and $dm_3=0$.
\par
Figure 2 shows the evolution of the bulk Lorentz factors of regions 2 and 3 by taking the initial isotropic
kinetic energy $E_{\rm K,2,0}=10^{51}\rm ~ergs$ and the initial Lorentz factor $\gamma_{2,0}=150$ of region 2 at
the deceleration radius, and $n_1=1~\rm cm^{-3}$. The shape of the curves is obviously consistent with the
analysis in D04, who points out that the evolution can be divided into three stages: (I) $\gamma_2\propto
t^{-3/8}, \gamma_3\propto t^{-39/136}$; (II) $\gamma_2\sim\gamma_3\propto t^{-1/4}$; (III) $\gamma_2\propto
t^{-3/8}, \gamma_3\propto t^{-7/16}$.
%
%
\begin{figure}
\resizebox{\hsize}{!}{\includegraphics{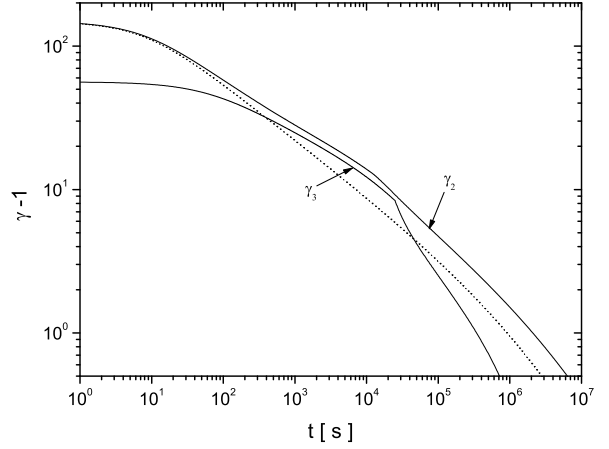}} \caption{The
evolution of $\gamma_2$ and $\gamma_3$ in the RWB model with
$B_{\perp,14}=2$ and $z=1$. The dotted curve represents the bulk
Lorentz factor of the shocked medium without wind injection. The
initial value of $\gamma_2$ is taken to be 150. This order of
magnitude has been suggested by some observations (e.g., Molinari et
al. 2006)}
\end{figure}
%
Compared with the case without wind injection, a significant change in the evolution of $\gamma_2$ is the
existence of stage II, which is quite similar to the result of the PEMI model (Wang \& Dai 2001). This
similarity can be easily understood since the term $4\pi R^2P_3dR={2\over3}(1+z)^{-1}L_{\rm w}dt$, which
represents the work done by region 3 to region 2 in Eq. 2, is approximately equal to (strictly, a factor $2/3$
times) the term $(1+z)^{-1}L_{\rm w}dt$ in the dynamic equation of the PEMI model for the same magnetar (for
details, see Dai \& Lu 1998a, 1998b; Wang \& Dai 2001; Fan \& Xu 2006). Therefore, we cannot expect to
distinguish the RWB model from the PEMI model only by observing the radiation from the forward-shocked medium. A
possible difference between these two energy-injection models is induced by reverse-shocked pairs, the radiation
of which arises from the remaining one-third energy release of the magnetar.

\section*{3. Radiation and example fit to X-ray afterglows}
Both the forward shock and the reverse shock heat cold materials to a higher temperature, generate random
magnetic fields, and accelerate protons and electrons. Since the microphysical processes have been unclear so
far, the electron energy density and the magnetic energy density are parameterized as usual. We assume that for
region 3 the electron and magnetic field energy densities are fractions $\epsilon_{\rm e,R}$ and $\epsilon_{\rm
B,R}$ of the total energy density behind the reverse shock ($\epsilon_{\rm e,R}+\epsilon_{\rm B,R}=1$), and for
region 2, fractions $\epsilon_{\rm e,F}$ and $\epsilon_{\rm B,F}$ of the total energy density behind the forward
shock, where $\epsilon_{\rm e,F}+\epsilon_{\rm B,F}<1$ and $\epsilon_{\rm e,F}\sim\sqrt{\epsilon_{\rm B,F}}$
according to Medvedev (2006). It is natural to think that $\epsilon_{\rm B,R}\neq\epsilon_{\rm B,F}$ and
$\epsilon_{\rm e,R}\neq\epsilon_{\rm e,F}$, as suggested in some studies (Fan et al. 2002; Coburn \& Boggs 2003;
Zhang, Kobayashi \& M\'esz\'aros 2003; Kumar \& Panaitescu 2003). We also assume that the spectral indices of
the electron energy distribution are $p_2$ and $p_3$ for regions 2 and 3, respectively. Furthermore, we assume
here that $p_2\sim p_3=p$. By giving a set of three free parameters ($p,~\epsilon_{\rm B,F}$ and $\epsilon_{\rm
B,R}$), we can fix the cooling Lorentz factors $\gamma_{\rm e,c}$ arising from both synchrotron radiation and
inverse-Compton scattering, and the minimum Lorentz factors $\gamma_{\rm e,min}$ in the electron energy
distributions of regions 2 and 3. We also consider the ``equal-arrival surface" effect of the RWB emission. The
related formulas for these calculations were presented in Sect. 3 of Huang et al. (2000). We ignore the
self-absorption effect for the X-ray emission we are interested in.
%
%
\begin{figure}
\resizebox{\hsize}{!}{\includegraphics{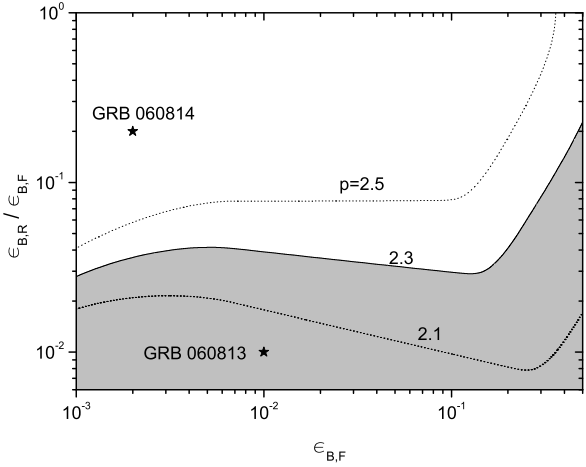}} \caption{Parameter
space of $(p,~\epsilon_{\rm B,F},~\epsilon_{\rm B,R})$ with
$B_{\perp,14}=2$ and $z=1$. The lines represent the critical values
$(\epsilon_{\rm B,R}/\epsilon_{\rm B,F})_{\rm c}$ with different
$p$, below which ( in the shaded area for $p=2.3$) the corresponding
reverse shock emission is insignificant for the total flux of the
RWB. The star symbols label the parameter sets of two GRBs.}
\end{figure}
%
\par
Before light curves are exhibited, it should be pointed out that the ratio of the flux contributed by regions 2
or 3 to the total flux is sensitive to the ratio of $\epsilon_{\rm B,R}/\epsilon_{\rm B,F}$. As shown in Fig. 3,
if a dot representing a parameter set $(p,~\epsilon_{\rm B,F},~\epsilon_{\rm B,R})$ is in the shaded area (for
$p=2.3$), the peak of the light curve of the reverse shock emission is below the light curve of the forward
shock emission (e.g., see the upper panel of Fig. 4). The higher the value of $p$, the easier the occurrence of
this case. Because the forward shock dominates the radiation of the RWB, as discussed above, the combined light
curve should be similar to the result of the PEMI model for an identical magnetar. In other words, all the GRB
X-ray afterglows that can be fitted in the PEMI model must be explained by the RWB model with an appropriate
parameter set. Conversely, if $\epsilon_{\rm B,R}/\epsilon_{\rm B,F}$ is large enough ($>(\epsilon_{\rm
B,R}/\epsilon_{\rm B,F})_{\rm c}$), the radiation from region 3 becomes quite important, especially during a
period of $\sim10^2-10^{5}\,$s. Therefore, we can divide the RWBs into two types: forward-shock-dominated and
reverse-shock-dominated bubbles.
\par
However, because of the existence of stage II in the dynamic evolution, the afterglows produced by two types of
RWBs always have a shallow decay phase. This feature of RWBs is very consistent with a lot of afterglows
observed by Swift. For example, two X-ray afterglows of GRB 060813 (upper panel) and GRB 060814 (lower panel)
are fitted by the RWB model as shown in Fig. 4, and the corresponding parameter sets $(p,~\epsilon_{\rm
B,F},~\epsilon_{\rm B,R})$ are shown in Fig. 3. It can clearly be suggested that the afterglows of GRB 060813
and GRB 060814 could be produced by forward-shock-dominated and reverse-shock-dominated RWBs, respectively. As
discussed above, the prompt emission of a GRB may result from internal shocks. If the high-latitude emission of
the last internal shocks is weaker than the early afterglow emission, one can observe the shallow decay of an
early afterglow, as in GRB 060813. Conversely, an early steep decay should be observed (Kumar \& Panaitescu
2000), as in GRB 060814. For the latter burst, we should thus introduce this curvature effect to explain the
earlier steep decay, which is fitted roughly by the following equation (as shown in the inserting panel of Fig.
4, also see Zhang et al. 2006; Liang et al. 2006):
\begin{equation}
F_{\rm X}(t)=A\left({t-t_{0}\over t_{0}}\right)^{-(2+\beta)}+F_{\rm
X, RWB},
\end{equation}
where $A$ is the normalization parameters, $t_{0}$ the time zero point of the last prompt emission pulse, and
$\beta$ the spectral index that is equal to 0.67 (Moretti, Guidorzi \& Romano 2006). Please note that we take
the time zero point of the afterglow as the GRB triggering time as proved by Kobayashi \& Zhang (2006) and Liang
et al. (2006).
%
%
\begin{figure}
\resizebox{\hsize}{!}{\includegraphics{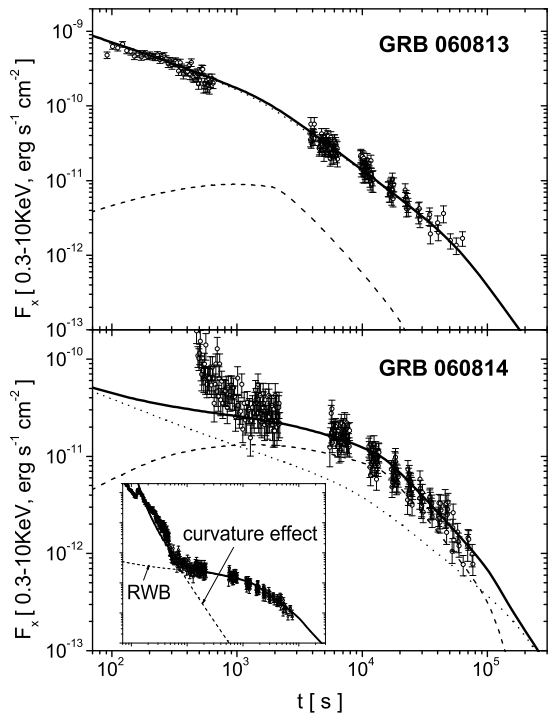}} \caption{Fits of GRB 060813 with $B_{\perp,14}=6.5, p=2.5,
\epsilon_{\rm B,F}=0.01, \epsilon_{\rm B,R}=0.0001, z=0.6$, and GRB 060814 with $B_{\perp,14}=2.5, p=2.2,
\epsilon_{\rm B,F}=0.002, \epsilon_{\rm B,R}=0.0004, z=0.6$ using the RWB model. The half-opening angles of jets
are $\theta_j=0.1$ and $\theta_j=0.15$, respectively. The solid lines correspond to total fluxes, while the
dotted and dashed lines represent the fluxes determined by the forward and reverse shocks, respectively. The
early steep decay segment of GRB 060814 is fitted with $A=9.85\times10^{-9}\rm ~erg~s^{-1}cm^{-2}$ and
$t_{0}=75~\rm s$ in the insert.}
\end{figure}
%
\section*{4. Summary}
In this paper, we have numerically calculated the dynamic evolution and radiation of RWBs. Two motivations impel
us to consider this RWB model rather than the PEMI model: first, the PEMI model cannot explain the very flat
plateau in light curves of some X-ray afterglows; second, a pulsar-driven energy outflow, which is dominated by
Poynting flux at smaller radii, could evolve into an ultrarelativistic electron-positron-pair wind at larger
radii. The most significant feature of the light curves of the X-ray afterglows from the RWBs is a flattening
segment occurring during a period of $\sim10^2-10^{5}\,$s, which is consistent with the observed shallow decay
phase. Our example fits to the X-ray afterglows of GRB 060813 and GRB 060814 indicate that they could be
produced by forward-shock-dominated and reverse-shock-dominated RWBs, respectively. This suggests that the
central engines of these two GRBs could be millisecond magnetars. Moreover, the example fits also show that,
besides the X-ray afterglows that can be fitted in the PEMI model (e.g., GRB 060813), the RWB model can also
explain some other afterglows with a very flat plateau (e.g., GRB 060814) that create difficulties for the PEMI
model.
\par
In addition, we would like to point out that the magnetar-driven RWB model is one possibility. As pointed out in
the introduction and in D04, RWBs could also be produced, in principle, by central black holes that accrete
circumburst materials after the GRB. Moreover, some other models have also been proposed to explain the observed
shallow decay phase, e.g., off-beam jets, two-component jets, and varying microphysics parameters (Eichler \&
Granot 2006; Granot et al. 2006; Panaitescu et al. 2006; also see Zhang 2007 for a recent review). It is a
demanding task to distinguish between the above models by using multiwavelength observations.

\begin{acknowledgements}
We would like to thank Binbin Zhang for providing the data of GRB 060813 and GRB 060814 and Bing Zhang for
valuable comments. YWY thanks Y.F. Huang for kind help and X.W. Liu for helpful discussions. This work is
supported by the National Natural Science Foundation of China (grant no. 10221001 and 10233010). YWY is also
supported by the Visiting Graduate Student Foundation of Nanjing University and the National Natural Science
Foundation of China (grant no. 10373007).
\end{acknowledgements}


\end{document}